\documentclass[11pt]{article} 
\usepackage{hyperref} 
\pdfoutput=1 
 
\begin{document} 
 
\title{Plasma-based Control of Supersonic Nozzle Flow} 
 
\author{Datta V. Gaitonde\\ 
\\\vspace{6pt} Computational Sciences Branch, AFRL/RBAC\\
Air Vehicles Directorate \\ Air Force Research Laboratory, WPAFB, OH 45433, USA} 
 
\maketitle 
 
 
\begin{abstract} 
The flow structure obtained when Localized Arc Filament Plasma
Actuators (LAFPA) are employed to control the flow issuing from a
perfectly expanded Mach 1.3 nozzle is elucidated by visualizing
coherent structures obtained from Implicit Large-Eddy Simulations.
The computations reproduce recent experimental observations at the
Ohio State University to influence the acoustic and mixing properties
of the jet. Eight actuators were placed on a collar around the
periphery of the nozzle exit and selectively excited to generate
various modes, including first and second mixed ($m = \pm 1$ and $m =
\pm 2$) and axisymmetric ($m = 0$).  In this fluid dynamics video (\href{http://ecommons.library.cornell.edu/bitstream/1813/13723/2/Alljoinedtotalwithmodetextlong2-Datta\%20MPEG-1.m1v}{mpeg-1}, \href{http://ecommons.library.cornell.edu/bitstream/1813/13723/3/Alljoinedtotalwithmodetextlong2-Datta\%20MPEG-2.m2v}{mpeg-2}),
unsteady and phase-averaged quantities are displayed to aid
understanding of the vortex dynamics associated with the $m = \pm 1$
and $m=0$ modes excited at the preferred column-mode frequency
(Strouhal number $0.3$).  The unsteady flow in both contains a broad
spectrum of coherent features.  For $m=\pm 1$, the phase-averaged flow
reveals the generation of successive distorted elliptic vortex rings
with axes in the flapping plane, but alternating on either side of the
jet axis. This generates a chain of structures where each interacts
with its predecessor on one side and its successor
on the other. Through self and mutual interaction, the leading segment
of each loop is pinched and passes through the previous ring before
rapidly breaking up, and the mean jet flow takes on an elliptic
shape. The $m = 0$ mode exhibits relatively stable roll-up events,
with vortex ribs in the braid regions connecting successive large
coherent structures.  Results with other modes are described in
Ref. 1.
\end{abstract} 
 
 
\section{Introduction} 
Jet flow behavior and control has significant consequences for many
applications, including mitigation of aircraft noise, and mixing for
various industrial concerns such as combustion and pollution.
Numerous types of control methods have been employed, including
passive (chevrons, tabs) and active (fluidic and plasma) techniques.
The present effort uses numerical simulations to explore the dynamics
of plasma-based open-loop flow control of a Mach 1.3 jet, described by
Samimy et al [2]. The plasma devices, denoted Localized Arc Filament
Plasma Actuators (LAFPA), strike an electric arc at a specified
frequency between two closely placed pin electrodes.  Their advantages
include rapid on-off capability, low inertia and superior high
frequency performance.  The full 3-D Navier-Stokes equations are
solved with an implicit LES approach. The effect of the actuators is
simulated with a surface heating model that successfully reproduces
the principal effects observed in experimental flow visualizations.
Since the actuators can be either on or off, the excitation is applied
in a discrete fashion and can be described by a collective frequency
and duty cycle.  The fluid dynamics video may be found in
\href{http://ecommons.library.cornell.edu/bitstream/1813/13723/2/Alljoinedtotalwithmodetextlong2-Datta\%20MPEG-1.m1v}{mpeg-1}
and 
\href{http://ecommons.library.cornell.edu/bitstream/1813/13723/3/Alljoinedtotalwithmodetextlong2-Datta\%20MPEG-2.m2v}{mpeg-2}
formats. The visualizations explore first mixed
(flapping) and axisymmetric modes at a Strouhal number of $0.3$,
corresponding to $4618Hz$ and a duty cycle of $20\%$.  The spreading
rate of the jet with the $m = \pm 1$ mode is significantly increased
in the flapping plane, but is reduced in the non flapping-plane
relative to the no-control case. Phase-averaged observations
visualized with isolevels of the Q-criterion colored by vorticity
magnitude, and vorticity magnitude colored by velocity magnitude,
indicate staggered and unstaggered structures on the two planes
respectively. The video shows these to be consistent with rings whose
axes lie in the flapping plane, but alternate about the jet
centerline. The resulting interaction between successive such events
results in elongation and partial pushing through of the leading
segment of each ring through the trailing segment of the previous ring.
The jet displays an elliptic cross-section in the mean. Axisymmetric
excitation yields successive roll-up events subjected to azimuthal
disturbances, whose rapid evolution leads to breakdown. Although not
shown in the video, the $m =\pm 2$ mode (see Ref. [1]) also generates
elliptic vortex structures but with major axes being successively
aligned along the two symmetry planes. The minor axes regions
move downstream faster because of the higher velocity near the
centerline, yielding rings which stretch in the streamwise direction
and ultimately breakdown.

\section*{References}
[1] Gaitonde, D.V., ``Simulation of Supersonic Nozzle Flows with Plasma-based Control,'' AIAA-2009-4187,
39th AIAA Fluid Dynamics Conference, San Antonio, Texas, June 22-25, 2009.\newline
[2]  Samimy, M., Kim, J.-H., Kastner, J., Adamovich, I., and Utkin, Y., ``Active Control of High-speed and High-Reynolds-number Jets Using 
Plasma Actuators,'' J. Fluid Mech., Vol. 578, 2007, pp. 305-330

\end{document}